\documentclass[nonatbib,nocopyrightspace]{sigplanconf}
\usepackage[numbers]{natbib}
\usepackage{inconsolata,alltt}
\usepackage[small,compact]{titlesec} 
\usepackage{enumitem}

\renewenvironment{thebibliography}[1]{%
   \begin{oldthebibliography}{#1}%
     \setlength{\itemsep}{-.3ex}%
}{%
   \end{oldthebibliography}%
}

\usepackage{amsmath}

\date{}
\begin{document}




\title{Taking Semantics Seriously in Program Analysis}

\authorinfo{Sam Tobin-Hochstadt \and David Van Horn}
           {PRL, Northeastern University}
           {}


\twocolumn[%
 \centerline{\LARGE \bf Semantic Solutions to Program Analysis Problems} 

 \medskip

 \centerline{\large Sam Tobin-Hochstadt \hspace{1cm}   David Van Horn}
 \smallskip
 \centerline{PRL, Northeastern University}
 \bigskip
 ]

\begin{abstract}
Problems in program analysis can be solved by developing novel program
semantics and deriving abstractions conventionally.  For over thirty
years, higher-order program analysis has been sold as a hard problem.
Its solutions have required ingenuity and complex models of
approximation.  We claim that this difficulty is 
due to premature focus on \emph{abstraction} and propose a new
approach that emphasizes \emph{semantics}.  Its simplicity enables new
analyses that are beyond the current state of the art.
\end{abstract}

\section*{Current Thoughts, New Ideas}

Higher-order program analysis has been an important and recurring
topic at PLDI, starting with Shivers' seminal paper
\cite{Shivers:1988:CFA:53990.54007} and continuing through the
present \cite{Might:2010:REK:1806596.1806631}.
However, past approaches are limited in the language features they can
handle, require intricate formal models that are difficult to develop,
verify, and maintain, and do not scale to new questions that we need
to answer of programs.  We propose a new approach in which interesting
\emph{analyses} can be developed by first developing interesting \emph{semantics}
and then using known techniques to approximate  as a final step.

As an example, Meunier, \emph{et
  al.}~\cite{Meunier:2006:MSA:1111037.1111057} develop a modular
program analysis for higher-order behavioral software contracts.
Meunier gives an analysis in the form of a large constraint set system
and, separately, a dynamic reduction semantics.  An important drawback
is the dissimilarity between the semantics and the analysis. Both are
complicated for the sake of establishing a correspondence, which is accomplished
by shoehorning the semantics into an analysis, and 
tweaking to achieve modularity.  Despite these efforts, the
soundness theorem does not hold.  Worse, the system was then
abandoned, as it could not be
maintained, extended, or implemented.

In contrast, we have taken the semantics of Meunier's language and
\emph{systematically derived} a similar whole-program analysis based
on an abstract machine for the language
\cite{DBLP:journals/corr/abs-1103-1362}.  The machine itself is
derived from the semantics through known techniques, making its
correctness proof straightforward.  This step is purely a semantic
refactoring; it has nothing to do with approximation.  The machine,
however, is in a form that abstracts naturally and transparently
\cite{VanHorn:2010:AAM:1863543.1863553}.

What remains is to make this analysis \emph{modular}, enabling
reasoning about programs that are missing some of their components.
We solve this problem purely on the
semantic side of the equation by extending the dynamic semantics with
reductions for programs with missing components.  Missing components
are regarded as their contracts, which are given reduction rules
corresponding to the reductions that may be taken by \emph{any} value
satisfying those contracts.  As an example, consider the following
program fragment consisting of two modules with unknown
implementations, {\tt keygen} and {\tt rsa}, and a call to {\tt rsa}
to encrypt a string using a key from {\tt keygen}.
Inputs and outputs are annotated with contracts, which are
user-defined predicates, i.e. {\tt prime?}.
\begin{alltt}
   keygen() : prime? \(\verb|{|\) \(\bullet\) \(\verb|}|\)
   rsa(k: prime?, s: string?) : string? \(\verb|{|\) \(\bullet\) \(\verb|}|\)
   rsa(keygen(), "Plain");
\end{alltt}
Under our modular semantics, the program executes as follows:
\[
\begin{array}{ll}
& \text{\tt{rsa(keygen(), "Plain");}}\\
\rightarrow& \text{\tt{rsa([prime?], "Plain");}}\\
\rightarrow&\text{\tt{string?("Plain"); prime?([prime?]); [string?]}}\\
\rightarrow&\text{\tt{[string?]}}
\end{array}
\]
The {\tt[\(\cdot\)]} notation denotes a contract treated as a value.
Intuitively, it represents the set of all values satisfying the
contract. The implementation of {\tt keygen} is missing, so we
cannot know what it returns, but by its specification, it produces a
value satisfying {\tt prime?}, hence it produces {\tt [prime?]}.  To
call {\tt rsa}, we check {\tt string?} of {\tt "Plain"} and {\tt
  prime?} of {\tt [prime?]}, both of which succeed, so the program produces
{\tt [string?]}, an unknown string value.  No contracts are
violated and thus expensive run-time checks can be eliminated.

To obtain an analysis, this \emph{modular} reduction semantics is
run through the  same derivation pipeline to reveal
a \emph{modular} program analysis.  The resulting analysis is easy to
verify, extend, and implement, requiring no ingenuity in
 approximation methods.

The central lesson of this work is that \emph{problems in program analysis can
  be solved by developing novel program semantics and deriving
  abstractions conventionally}.
Generalizing this observation, we can see that this strategy applies
to many analysis problems.
Determine the question to be answered,
 design a semantics that precisely answers this question during
evaluation, as in our modular semantics, and finally, use traditional  transformations and
approximation methods to produce a computable analyzer.

This strategy has several advantages: (1) It is easier to get right.
Semantics and analysis correspond closely, making both easier to
verify and maintain.  (2) Many existing semantics can be repurposed
for building analyses of everything from space behavior of lazy
languages to security via stack inspection.  (3) The PL community has
developed a host of intellectual tools for designing and reasoning
about semantics which we can re-use for program analysis.

\section*{Future Fun}

We have taken this approach to leverage dynamic semantics for
predicting garbage collection, space consumption, and modularity.
There are many exciting opportunities we consider worth pursuing.

\begin{enumerate}[topsep=1pt, partopsep=1pt]
\setlength{\itemsep}{0pt}
     \setlength{\parskip}{0pt}
\item Using a parallel cost model semantics~\cite{citeulike:9217818},
  we can design analyses for predicting space usage of parallel
  functional programs.

\item Contracts are a form of specification, which we treat as
  values in a novel semantics.  What other kinds of specifications can
  be treated as values?  Giving reductions for values drawn from
  Hoare-type theory~\cite{Nanevski:2008:HTT:1520009.1520017} would
  give rich specifications for effectful components, in turn yielding
  rich program analyzers.

\item Using a semantics with temporal predicates over program events~\cite{Skalka:2008:TTE:1348945.1348947}, we can develop
  higher-order temporal model checkers for history- and stack-based
  security mechanisms.
\end{enumerate}
These problems seem daunting under current approaches to analysis
design, but we conjecture that by taking a semantic approach
seriously, solutions will be more easily obtained.

\vskip 4pt
\noindent \emph{\bfseries Acknowledgments:}
We are inspired in part by work with M. Might.

{\scriptsize
  \bibliographystyle{unsrt}
  \bibliography{bibliography}
}

\end{document}